\pdfoutput=1
\documentclass[11pt]{article}

\usepackage[margin=1in,headheight=14pt]{geometry}
\usepackage[T1]{fontenc}
\usepackage[scale=0.88]{inter}
\usepackage{libertinus-type1}
\usepackage{libertinust1math}

\usepackage[scaled=0.95]{inconsolata}
\usepackage[final,kerning=true]{microtype}
\SetExtraKerning[unit=1em]{encoding=T1,family=Inter-TLF}{-={-90,-90}}
\usepackage[table]{xcolor}

\definecolor{dsblue}{HTML}{4D6BFE}
\definecolor{dslink}{HTML}{3550DC}
\definecolor{dsnavy}{HTML}{1B2559}
\definecolor{dsink}{HTML}{23272F}
\definecolor{dsgrey}{HTML}{6B7280}
\definecolor{dsrule}{HTML}{C9D1F5}
\definecolor{dstint}{HTML}{F3F5FF}
\color{dsink}

\newif\ifanon
\anonfalse

\usepackage{hyperref}
\ifanon
\hypersetup{colorlinks=true, linkcolor=dslink, citecolor=dslink, urlcolor=dslink,
  pdftitle={A Two-Count Baseline for Change Detection on Dynamic Networks}, pdfauthor={}}
\else
\hypersetup{colorlinks=true, linkcolor=dslink, citecolor=dslink, urlcolor=dslink,
  pdftitle={A Two-Count Baseline for Change Detection on Dynamic Networks},
  pdfauthor={Izhar Ali}}
\fi
\usepackage{url}
\usepackage{graphicx}
\graphicspath{{figures/}}
\usepackage{booktabs}
\usepackage{colortbl}
\arrayrulecolor{dsnavy}
\let\dsmidrule\midrule
\renewcommand{\midrule}{\arrayrulecolor{dsrule}\dsmidrule\arrayrulecolor{dsnavy}}
\let\dscmidrule\cmidrule
\renewcommand{\cmidrule}{\arrayrulecolor{dsrule}\dscmidrule}
\usepackage{enumitem}
\usepackage{amsmath}
\usepackage{amssymb}
\usepackage{mathtools}
\usepackage{amsthm}
\usepackage[round,authoryear]{natbib}
\usepackage{doi}

\usepackage[skins,breakable]{tcolorbox}

\usepackage[explicit]{titlesec}
\titleformat{\section}{\sffamily\Large\bfseries\color{dsnavy}}
  {\textcolor{dsblue}{\thesection}}{0.7em}{#1}
\titleformat{\subsection}{\sffamily\large\bfseries\color{dsnavy}}
  {\textcolor{dsblue}{\thesubsection}}{0.6em}{#1}
\titleformat{\paragraph}[runin]{\sffamily\normalsize\bfseries\color{dsnavy}}{}{0pt}{#1}
\titlespacing*{\section}{0pt}{3.0ex plus 1ex minus .3ex}{1.4ex plus .3ex}
\titlespacing*{\subsection}{0pt}{2.2ex plus .8ex minus .2ex}{1.0ex plus .2ex}
\titlespacing*{\paragraph}{0pt}{1.3ex plus .5ex minus .2ex}{0.8em}

\usepackage[font=small,labelfont={sf,bf,color=dsblue},labelsep=period,skip=6pt]{caption}
\usepackage{fancyhdr}
\fancypagestyle{plain}{\fancyhf{}\fancyfoot[C]{\sffamily\footnotesize\color{dsgrey}\thepage}}

\usepackage[capitalize,noabbrev]{cleveref}
\crefname{appendix}{Appendix}{Appendices}

\newtheoremstyle{dsplain}{0pt}{0pt}{\itshape}{}{\sffamily\bfseries\color{dsnavy}}{.}{0.5em}
  {\thmname{#1}\thmnumber{ #2}\thmnote{ \mdseries(#3)}}
\theoremstyle{dsplain}

\theoremstyle{definition}

\tcbset{dsthm/.style={enhanced, breakable, colback=dstint, colframe=dstint, boxrule=0pt, arc=1.5pt,
  borderline west={2.4pt}{0pt}{dsblue}, left=9pt, right=8pt, top=5pt, bottom=5pt, before skip=10pt, after skip=10pt}}
\tcolorboxenvironment{theorem}{dsthm}
\tcolorboxenvironment{proposition}{dsthm}
\tcolorboxenvironment{lemma}{dsthm}

\newcommand{\dstitle}{%
  \thispagestyle{plain}%
  \begin{flushleft}
    {\sffamily\fontsize{19.5}{25}\selectfont\bfseries\color{dsnavy}A Two-Count Baseline for\\[2pt]
     \color{dsblue}Change Detection on Dynamic Networks\par}
    \vspace{12pt}
    \ifanon
    {\sffamily\normalsize\bfseries Anonymous authors}\par\vspace{2pt}
    {\sffamily\small\color{dsgrey}Paper under double-blind review}\par
    \else
    {\sffamily\normalsize\bfseries Izhar Ali}\par\vspace{2pt}
    {\sffamily\small\color{dsgrey}Department of Computer Science, Rowan University\quad
     \href{mailto:aliz@rowan.edu}{aliz@rowan.edu}\\
     Code: \href{https://github.com/ali-izhar/two-count-baseline}{github.com/ali-izhar/two-count-baseline}}\par
    \fi
    \vspace{6pt}
  \end{flushleft}\vspace{2pt}}
\renewenvironment{abstract}{%
  \begin{tcolorbox}[enhanced, colback=dstint, colframe=dstint, boxrule=0pt, arc=1.5pt, left=12pt, right=12pt,
    top=9pt, bottom=9pt, borderline west={2.4pt}{0pt}{dsblue}]
  {\sffamily\small\bfseries\color{dsnavy}Abstract}\par\vspace{4pt}\small}{\end{tcolorbox}\vspace{4pt}}

\begin{document}
\dstitle

\begin{abstract}
Change-point detectors for dynamic networks aim to find changes in how a network is connected. They are
evaluated on real networks with labelled events, but many of these events also change how active the
network is: how many vertices take part and how many edges are present. A detector can therefore score
well by responding to activity alone, without detecting any change in structure. We test whether detectors 
from the literature beat a simple baseline that tracks only the number of active vertices and the number of 
edges in each snapshot, a baseline that we motivate with a standard statistical test for a change in 
activity. On ten datasets, four of them from the papers of published detectors, none of thirteen detectors 
ranks the labelled events significantly better than the baseline when the results are pooled (AUC $0.75$, 
or $0.73$ with scales taken from early snapshots only, against $0.66$ for the best detector). The baseline is best on only two datasets, however, and on the four 
datasets with the most events three detectors that track individual vertices score higher. When we rewire 
networks while keeping the degree of every vertex fixed, most detectors that ignore vertex identities barely 
respond, but several that track individual vertices do. New detectors should be compared with this baseline 
and tested at fixed degree.
\end{abstract}

\section{Introduction}
\label{sec:intro}

The MIT Social Evolution study recorded, day by day, which students of one dormitory were near one
another \citep{madan2012sensing}. Each day gives a graph, or \emph{snapshot}, whose edges join the
students who were in contact, and the days form a series. Over Christmas 2008 the number of students
active per day, those in at least one contact, fell from $34$--$43$ to $3$--$8$. A change-point detector
scores each day by how much the recent snapshots differ from earlier ones, and it is judged on labelled
events such as Christmas: it should score the days after the event above every stretch of the same length
that contains no event, an \emph{event-free window}. LAD \citep{huang2020laplacian}, a detector built on
the Laplacian spectrum, meets this standard at Christmas, and so does a score made of two counts: the
number of active students and the number of pairs in contact. LAD's alarm is correct, but it cannot show 
that the pattern of contacts changed, because both counts changed with it.

Published detectors summarize each snapshot by a quantity that depends on its structure: the Laplacian
eigenvalues (LAD), the distribution of the eigenvalues, or density of states (SCPD,
\citealp{huang2023fast}), or a similarity learned by a graph neural network (s-GNN,
\citealp{sulem2023graph}). \emph{Permutation-invariant} detectors give the same score when the vertices
are relabelled; \emph{node-aligned} detectors compare the same vertices across snapshots. Either kind can
respond to how many vertices and edges are present as well as to how they are connected.

\citet{kendrick2018change} found on weekly snapshots of two networks, MIT Reality Mining and Enron, that the
number of active vertices and the number of edges, each fed to a Bayesian change-point method, detect the
labelled changes nearly as well as summaries of fitted block and hierarchical models. The detectors
published since (LAD, SCPD, s-GNN and MultiLAD, \citealp{huang2024laplacian}) do not compare against these
counts, and their benchmarks have further weaknesses. On AskUbuntu, whose events are Ubuntu
releases six months apart, a score that never changes reaches $F_1 = 1.00$ when each alarm is followed by
a fixed quiet period: the alarms then recur at even intervals, as the events do. On MAG, and on two benchmarks 
that we cannot score, the labelled events are dates that the detector introducing the benchmark flagged and 
that its authors then matched to historical events. Last, we could not reproduce the reported results of s-GNN 
with its released code.

This paper asks what the benchmark events of these detectors change, and what the detectors see when
activity is held fixed. We score thirteen detectors from the literature and two counts on the benchmarks of LAD,
SCPD and s-GNN and on six other datasets, ranking each event against the event-free windows of its series.
\begin{enumerate}[leftmargin=*,itemsep=1pt,topsep=2pt]
  \item A diagnosis of the benchmarks. Labelled events are large changes of activity far more often than
  event-free windows (an exploratory comparison), and pooled over ten datasets no detector ranks them
  significantly better than the two counts (\cref{sec:events}).
  \item A reason to use the counts as the control. A change of a graph splits into a change of activity
  and a change of structure, and in a Rasch model the window change of each count is the conditional score
  statistic for a change of activity (\cref{sec:control}).
  \item Tests at fixed degree, by rewiring real series and by block models of matched expected degree. Most
  permutation-invariant detectors barely respond to rewiring, while degree assortativity and several
  node-aligned detectors do (\cref{sec:structure}).
  \item A protocol for evaluating a new detector of structural change (\cref{sec:discussion}).
\end{enumerate}
Compared with \citet{kendrick2018change}, we evaluate the detectors that claim to see structure on their
own benchmarks, rank events instead of thresholding scores, motivate the counts as the control, and
test what the detectors see at fixed degree.

\section{Related work}
\label{sec:related}

\paragraph{Activity as a confound.} Many network statistics depend on the number of edges, so a change of
that number can trigger or hide a detection \citep{lafond2014anomaly,lafond2018designing}, as in brain
networks \citep{vanwijk2010comparing}. \citet{peel2015detecting} compared their generalized hierarchical
random graph (GHRG) detector with mean-degree and clustering baselines; the size-agnostic detector of
\citet{miller2020size} remains sensitive to density by its authors' account, and surveys of graph
distances report their dependence on size and density
\citep{donnat2018tracking,wills2020metrics,hartle2020network}.

\paragraph{Trivial baselines and metrics.} One line of code solves many time-series anomaly benchmarks
\citep{wu2021current}, and a method that never reports a change beats many change-point methods
\citep{vandenburg2020evaluation}. In link prediction and graph classification, memorization and graph-free
baselines compete with graph models \citep{poursafaei2022towards,errica2020fair}. Point-adjusted $F_1$
rates a random score highly \citep{kim2022towards}, and range-based and volume-under-surface scores
\citep{tatbul2018precision,paparrizos2022volume} grade a whole series at once. We rank each event on its
own and include a constant score.

\paragraph{Detectors we do not run.} Model-based detectors fit a GHRG \citep{peel2015detecting} or
monitor the block densities of a degree-corrected block model \citep{wilson2019modeling}, and a martingale
detector tests the exchangeability of graph features \citep{ho2026detecting}; none is run on our datasets
(\cref{app:settings}). Statistical work on correlation networks
\citep{barnett2016change,cribben2017estimating}, graphons \citep{zhao2019change}, missing links
\citep{enikeeva2025change} and online detection \citep{yu2024network} compares procedures with one another
or with scan statistics.

\section{Activity, structure and the counts}
\label{sec:control}

\paragraph{What a change can alter.} A snapshot $G_t$ is an undirected graph on a vertex set of size $n$,
and a vertex is \emph{active} if it has an edge. A change can alter four things: (i)~the numbers of active
vertices $n_+$ and of edges $|E|$; (ii)~\emph{membership}, which vertices are active; (iii)~the
\emph{labelled structure}, which labelled pairs are connected, given (i) and (ii); and (iv)~the
\emph{unlabelled structure}, the properties that do not depend on the vertex labels and are not fixed by
the degrees, such as the triangles, the degree assortativity \citep{newman2002assortative} or the blocks.
\emph{Activity} is (i) and (ii); \emph{structure} is (iii) and (iv). Two students who swap all their contacts alter (iii) only. A family of scores
can see a change if, for some graphs, the change alters at least one score of the family. A score built on
$n_+$ and $|E|$ sees (i) only, and \emph{vertex turnover}, the number of vertices that become active or
inactive, sees (i) and (ii); neither can respond to structure, whatever the data. \Cref{tab:sees} shows
which change each family can see, and where we test it.

\begin{table}[ht]
  \centering
  \caption{Which change each family of scores can see (a tick). Perm.-inv.: the
  permutation-invariant detectors and statistics such as transitivity and degree assortativity.}
  \label{tab:sees}
  \footnotesize\setlength{\tabcolsep}{3pt}
\begin{tabular}{llccccl}
  \toprule
  Change & Alters & Counts & Turnover & Perm.-inv. & Node-aligned & Tested in \\
  \midrule
  A labelled event, e.g.\ Christmas at MIT & (i), (ii), possibly (iii), (iv) & \checkmark & \checkmark & \checkmark & \checkmark & \cref{sec:events} \\
  The same graph on other active vertices & (ii), (iii) & & \checkmark & & \checkmark & \cref{sec:events} \\
  Degree-preserving rewiring & (iii), (iv) & & & \checkmark & \checkmark & \cref{sec:structure} \\
  Relabelling of the active vertices & (iii) & & & & \checkmark & \cref{sec:structure} \\
  \bottomrule
\end{tabular}
\vspace{-8pt}
\end{table}

\paragraph{The counts.} A window-based score compares a test window, the $w$ snapshots before step $t$,
with a reference window, the $w_{\mathrm{ref}}$ snapshots before those. We divide $n_+$ and $|E|$ by their
scales, the median absolute change between consecutive snapshots, and write $\Delta_{n_+}$ and
$\Delta_{|E|}$ for the changes of the window means of the two scaled numbers. The score at $t$ is
\[
  s_t = \bigl((\Delta_{n_+})^2 + \lambda\,(\Delta_{|E|})^2\bigr)^{1/2},
\]
with $\lambda = 1/4$, $w = 2$ and $w_{\mathrm{ref}} = 5$, chosen as in \cref{sec:protocol}; the weights
$\lambda = 1/4$, $1/2$ and $1$ give pooled EP within $0.002$ of one another (\cref{app:settings}). We call
this baseline score \emph{the counts}.

A series may consist of \emph{segments}, contiguous recordings such as one day of contacts or one
subject's sensor data. Scales taken over the whole series use future snapshots, so the counts are offline;
the \emph{causal counts} take each scale from the first $20$ snapshots of each segment only. We report both.

\paragraph{The counts and the score test.} Suppose that a change of activity adds the same amount $\beta$
to the log-odds that each unit (a vertex or a vertex pair) is present, while each unit keeps its own
baseline rate. This is the Rasch model \citep{rasch1960probabilistic} with a window effect: unit $i$ is
present in snapshot $s$ with log-odds $\alpha_i + \beta\,\mathbf 1[s \text{ in the test window}]$,
independently across units and snapshots given the parameters. Let $x_i$ and
$c_i$ count the test snapshots and all $N = w + w_{\mathrm{ref}}$ snapshots in which unit $i$ is present.
Conditioning on $c_i$ removes every $\alpha_i$ \citep{cox1958two,andersen1970asymptotic}, and the
conditional score statistic for $\beta = 0$ is the Mantel--Haenszel numerator $\sum_i (x_i - w c_i/N)$
\citep{mantel1959statistical}. Counting the present units snapshot by snapshot gives, for any 0/1 presence
array,
\begin{equation}
  \label{eq:score}
  \sum_i \Bigl(x_i - \frac{w\,c_i}{N}\Bigr) = \frac{w\,w_{\mathrm{ref}}}{N}\,
  \bigl(\bar n_{\mathrm{test}} - \bar n_{\mathrm{ref}}\bigr),
\end{equation}
where $\bar n_{\mathrm{test}}$ and $\bar n_{\mathrm{ref}}$ are the mean numbers of present units per
snapshot in the two windows. With vertices as units the right side is proportional to
the window change of $n_+$; with pairs, to that of $|E|$. Each term of the counts is thus, up to a
constant factor within a segment, the square of the score statistic for a change of activity at one level,
with every unit's own rate conditioned away. The identity
fixes the two terms but not their weights, so the counts are not the score test itself. Pooled, the
standardized pair-level score test, which divides by the conditional variance, ranks the labelled events of
\cref{sec:bench} as well as the counts, though not on every dataset; \cref{app:proofs} gives the proof and these results.

\section{Benchmark and protocol}
\label{sec:protocol}

\paragraph{Datasets.} \Cref{tab:data} lists the ten datasets. Four come from the papers of LAD, SCPD and s-GNN
and are built with their loaders and events: UCI, MAG, the stablecoin network and PAMAP2, the
\emph{source datasets}; on MAG the events are SCPD's own detections. Three more datasets of these papers, the
Senate, the Canadian bills and the flights, have all their events in the first $17$ snapshots, which no
method scores (\cref{app:data}). We built MIT and Enron with our own event lists, MIT from the
Social Evolution study with dates from the academic calendar \citep{mit2008calendar}, and added four public
datasets for their size and domain: the school, HT09, EEG and AskUbuntu. PAMAP2 and
EEG are correlation networks, in which an edge joins two channels whose signals are correlated above a
threshold; there $|E|$ measures how strongly the signals move together, and its change can be the event
itself, as when a seizure starts. Every snapshot is a binary symmetric graph on a fixed vertex set; on AskUbuntu, 
each month's users are mapped to fixed user ids for the methods that need them.

\begin{table}[ht]
  \centering
  \caption{The datasets. $T$: snapshots; $n$: vertices of the fixed vertex set (AskUbuntu: users per month, at most); 
  Events: labelled, with the number scored after the burn-in in parentheses; $^*$: the source detector's own detections;
  Event-free: event-free windows; $\tau$: tolerance in steps. Sources: LAD \citep{huang2020laplacian}, SCPD
  \citep{huang2023fast}, s-GNN \citep{sulem2023graph}; ours: our construction and event list; added: a public
  dataset that we added.}
  \label{tab:data}
  \footnotesize
  \setlength{\tabcolsep}{3.4pt}
  \begin{tabular}{lllrrlrr}
    \toprule
    Dataset & Source & Snapshot, events & $T$ & $n$ & Events & Event-free & $\tau$ \\
    \midrule
    UCI messages & LAD & day, term ends and starts & 196 & 1{,}900 & 2 (2) & 177 & 0 \\
    MAG history & SCPD & year, wars, a regime change & 181 & 502{,}549 & 5$^*$ (5) & 139 & 2 \\
    Stablecoin & SCPD & day, market news & 215 & 6{,}715{,}034 & 31 (30) & 85 & 3 \\
    PAMAP2 & s-GNN & 1\,s epoch, activity changes & 19{,}261 & 27 & 98 (97) & 18{,}016 & 5 \\
    MIT Social Evolution & ours & day, academic calendar & 270 & 63 & 6 (6) & 189 & 5 \\
    Enron, all addresses & ours & week, corporate events & 200 & 86{,}664 & 4 (4) & 147 & 4 \\
    Primary school & added & 5\,min, lunch & 189 & 242 & 4 (4) & 131 & 2 \\
    HT09 conference & added & 5\,min, sessions and breaks & 434 & 113 & 22 (22) & 283 & 2 \\
    EEG (CHB-MIT) & added & 2\,s epoch, seizure onsets & 33{,}562 & 21 & 19 (19) & 32{,}935 & 5 \\
    AskUbuntu & added & month, releases & 76 & $\le$\,8{,}102 & 11 (9) & 10 & 2 \\
    \bottomrule
  \end{tabular}
\end{table}

\paragraph{Event percentile.} The constant score on AskUbuntu shows that $F_1$ at a tuned threshold
rewards the alarm schedule as much as the score, so we score the ranking. Every method is scored from the
same step, after a \emph{burn-in} of the first $17$ snapshots of each segment, the longest span of windows any
method needs (LAD compares windows of $7$ and $10$ snapshots). An event at step $e$, its first changed snapshot, 
has the window $[e, e+\tau]$, where $\tau$ is the tolerance of its dataset; an event-free window is a run 
$[s, s+\tau]$ of scored steps within one segment that overlaps no event's window. The \emph{event percentile} 
(EP) of an event is the fraction of event-free windows of its series in which a method's highest score is 
lower than in the event's window, with ties counted as one half: the AUC of the event window against the 
event-free windows. The constant score has EP $0.5$. We average EP over the events of a dataset and then 
over datasets with equal weight, the \emph{pooled EP}, so that PAMAP2's $97$ events do not decide the result.

\paragraph{Detectors.} We compare the counts with thirteen detectors and three structural statistics, the sixteen methods. 
Six detectors are permutation-invariant: LAD, SCPD, and four general
change-point detectors that we run on eight summaries of each snapshot, among them $|E|$ and the mean
degree. The four are the autoencoder TIRE \citep{de2021autoencoder}, which is trained on the whole series 
and is therefore offline, the cumulative-sum test CUSUM \citep{page1954continuous}, Bayesian online 
change-point detection (BOCPD, \citealp{adams2007bayesian}) and maximum mean discrepancy 
(MMD, \citealp{gretton2012kernel}). Seven detectors are node-aligned, each adapted from its published 
procedure to compare a test window with a reference window: Hamming and weighted Jaccard distances, 
DeltaCon \citep{koutra2013deltacon}, the Chen--Zhang statistic \citep{chen2015graph}, network binary 
segmentation (NBS, \citealp{wang2021optimal}), adjacency spectral embeddings (ASE,
\citealp{sussman2012consistent,padilla2022change}) and a SpotLight-style sketch
\citep{eswaran2018spotlight}. The \emph{structural statistics} are the window changes of transitivity, of
degree assortativity and the larger of the two; the triangle count grows with the number of edges, so
transitivity also responds to activity. 

\paragraph{Reproduction.} LAD's code reproduces its reported results. SCPD's code recovers its detections on
MAG and the flights; on the stablecoin network six of its eight reported detections fall one day later and
none on the reported day. s-GNN's code reaches a mean point-adjusted $F_1$ of $0.14$ on PAMAP2, for which its paper gives no
settings, against $0.78$ reported, so we
leave s-GNN out (\cref{app:data}).

\paragraph{What was fixed when.} The settings of the counts and of CUSUM, BOCPD, MMD, LAD and TIRE were
chosen by the highest mean $F_1$ on MIT, Enron and AskUbuntu before any EP was computed, so the other
seven datasets are held out from that choice (\cref{app:settings}). The tolerances were fixed before any EP was
computed, and the two pools (all ten datasets, and the four with at least ten events) and the equal weight
per dataset were fixed at the first evaluation; the school, HT09 and EEG data and their events were fixed in a written protocol before any score
based on the counts was computed on them. The causal counts, the node-aligned detectors and the structural
statistics were added after the counts' first results. Analyses chosen after their own results are
exploratory and marked $^\circ$. The analysis specifications released with the code record these choices
and the reading rules used below.

\paragraph{Inference.} Each comparison of the counts with a method is tested twice, both times with Holm's
correction \citep{holm1979simple} over the sixteen methods: by an exact sign-flip test over the
per-dataset differences, and by a paired two-stage bootstrap that resamples datasets, then events, and
gives the intervals. A $t$ test over the per-dataset differences is
reported beside them. We invert the two dataset-level tests, sign-flip and $t$, with
Bonferroni's correction to bound how far any method's pooled EP can exceed that of the counts, and quote the
larger of the two bounds. On the four datasets with at least ten events no
sign-flip test can give $p$ below $0.125$, so there the bootstrap decides. With ten datasets the comparison
has little power: an
unadjusted test detects a lead of $0.12$ to $0.25$ in pooled EP with $80\%$ power, depending on the method
(\cref{app:inference}).

\begin{figure}[t]
  \centering
  \includegraphics[width=0.96\textwidth]{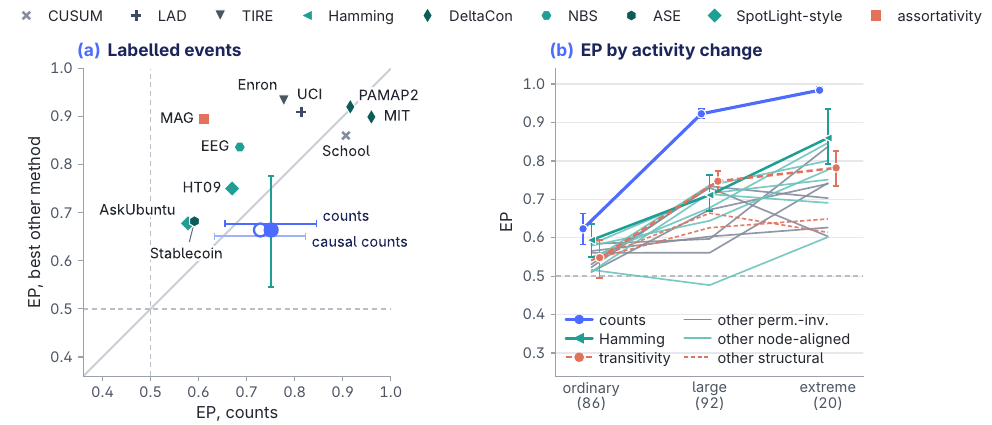}
  \caption{Labelled events. (a) EP of the counts (horizontal) and of the best of the sixteen methods on
  each dataset (vertical; the marker shows the method); choosing the best method after the fact favours it.
  Circles: pooled EP of the counts (filled) and of the causal counts (open) against that of Hamming, the best detector when pooled, with two-stage
  $95\%$ intervals. (b) Pooled EP by the activity change at the event (\cref{sec:events}); events in
  parentheses. The counts define these strata, so their rise is partly by construction.}
  \label{fig:main}
\end{figure}

\section{Results}
\label{sec:bench}

\subsection{Labelled events}
\label{sec:events}

\paragraph{Labelled events are large changes of activity three times as often as event-free windows.}
Averaged over the ten datasets, $45\%$ of the events are large or extreme changes of activity$^\circ$,
against $15\%$ of the event-free windows. To classify an event, we rank the window change of each count at
the event among its changes over the event-free windows of the series, and call the change \emph{large}
when the larger of the two percentiles lies between $90$ and $99$, \emph{extreme} above $99$, and
\emph{ordinary} otherwise; an event-free window is ranked the same way. The share among events is higher on
nine of the ten datasets; on MAG one of five events is large, against $20\%$ of the event-free windows
(\cref{app:inference}). Pooled over the datasets, every method ranks
the extreme events above the ordinary ones (\cref{fig:main}(b)), and within single datasets every method
except SCPD does so on average$^\circ$. On the extreme events the counts have the highest EP, partly by
construction.

\paragraph{Pooled, no detector ranks the events significantly better than the counts.} Pooled over the
ten datasets the counts
have EP $0.751$ and the causal counts $0.728$; the best detector, Hamming, has $0.664$, close to turnover and
transitivity (\cref{tab:pooled}). After Holm's correction the counts are significantly above BOCPD by both
tests and above MMD and SCPD by the bootstrap; no method is significantly above either version of the counts. The simultaneous
bounds limit any method's lead to $0.13$ over the counts and $0.18$ over the causal counts. The counts are at or
above every method only on MIT and the school network; on each of the other eight datasets some method is
above them, but seven different methods take those places
(\cref{fig:main}(a)). The counts keep the highest pooled EP under each check of \cref{app:inference}: without any one dataset, on
the seven held-out datasets, without events taken from detectors' own detections, with the stablecoin events
re-dated against the earliest public report$^\circ$, and at common tolerances$^\circ$ from $1$ to $10$ (without AskUbuntu from $\tau = 3$ on). Only at a tolerance of zero, and with every event moved two steps
earlier$^\circ$, is another method ahead: TIRE, which is offline, and not significantly.

\begin{table}[t]
  \centering
  \caption{Pooled EP, equal weight per dataset (datasets; events). Activity by design$^\circ$: events whose
  description names a change of activity; Other: the remaining $161$ events; $\ge 10$ events: HT09,
  stablecoin, PAMAP2 and EEG. Counts $-$ method: paired two-stage $95\%$ interval, unadjusted; positive: the
  counts are above. After Holm's correction over the sixteen methods, $\dagger$: significant by both tests;
  $\ddagger$: by the bootstrap only. Rows: references, permutation-invariant detectors, node-aligned
  detectors, structural statistics, each by descending EP on all ten. Bold: the highest detector or structural
  statistic per EP column. Differences from the causal counts on $\ge 10$ events: \cref{tab:pooledfull}.}
  \label{tab:pooled}
  \scriptsize
  \setlength{\tabcolsep}{3.2pt}
  \begin{tabular}{lccccc}
    \toprule
    & \multicolumn{2}{c}{All ten datasets} & Activity by design$^\circ$ & Other events & $\ge 10$ events \\
    \cmidrule(lr){2-3}
    Method & EP & Counts $-$ method & ($5$; $37$ events) & ($6$; $161$) & ($4$; $168$) \\
    \midrule
    Counts & 0.751 & -- & 0.827 & 0.690 & 0.716 \\
    Causal counts & 0.728 & $+0.022$ [$-0.016$, 0.068] & 0.795 & 0.680 & 0.645 \\
    Constant & 0.500 & $+0.251$ [0.155, 0.345] & 0.500 & 0.500 & 0.500 \\
    \midrule
    TIRE (offline) & 0.645 & $+0.106$ [$-0.021$, 0.246] & 0.552 & 0.694 & 0.671 \\
    LAD & 0.610 & $+0.141$ [0.009, 0.271] & 0.657 & 0.591 & 0.566 \\
    MMD & 0.604 & $+0.147$ [0.064, 0.227]$^\ddagger$ & \textbf{0.695} & 0.521 & 0.587 \\
    BOCPD & 0.587 & $+0.163$ [0.045, 0.307]$^\dagger$ & 0.603 & 0.575 & 0.682 \\
    CUSUM & 0.573 & $+0.178$ [0.037, 0.339] & 0.513 & 0.622 & 0.592 \\
    SCPD & 0.562 & $+0.189$ [0.069, 0.329]$^\ddagger$ & 0.566 & 0.571 & 0.683 \\
    \midrule
    Hamming & \textbf{0.664} & $+0.087$ [$-0.025$, 0.222] & 0.623 & 0.708 & 0.743 \\
    NBS & 0.656 & $+0.095$ [$-0.026$, 0.226] & 0.568 & \textbf{0.713} & 0.730 \\
    DeltaCon & 0.651 & $+0.099$ [$-0.023$, 0.237] & 0.570 & 0.642 & 0.608 \\
    Jaccard & 0.627 & $+0.124$ [0.008, 0.243] & 0.610 & 0.600 & 0.611 \\
    ASE & 0.615 & $+0.136$ [0.001, 0.296] & 0.550 & 0.683 & \textbf{0.749} \\
    SpotLight-style & 0.608 & $+0.142$ [0.002, 0.304] & 0.661 & 0.586 & 0.664 \\
    Chen--Zhang & 0.511 & $+0.240$ [0.068, 0.423] & 0.439 & 0.581 & 0.638 \\
    \midrule
    Transitivity & 0.661 & $+0.090$ [0.003, 0.175] & 0.605 & 0.685 & 0.635 \\
    Max.\ of the two & 0.597 & $+0.154$ [0.021, 0.274] & 0.577 & 0.571 & 0.540 \\
    Assortativity & 0.590 & $+0.161$ [0.024, 0.285] & 0.619 & 0.549 & 0.488 \\
    \midrule
    Turnover$^\circ$ & 0.665 & $+0.086$ [$-0.002$, 0.177] & 0.706 & 0.639 & 0.679 \\
    \bottomrule
  \end{tabular}
\end{table}

\paragraph{The lead is concentrated in events designed to change activity.} We call an event \emph{activity by
design}$^\circ$ when its description names a scheduled break, holiday, meal or session, the start or end of
a term, or a change in who takes part or in trading volume; the $37$ such events are all those of MIT, the
school, HT09 and UCI and three of the stablecoin network's. On them the counts have EP $0.827$ and no
detector or structural statistic has more than $0.695$. On the $161$ other events the counts have $0.690$, and
NBS ($0.713$), Hamming ($0.708$) and TIRE ($0.694$) are above them, none significantly.

\paragraph{On the datasets with the most events, three node-aligned detectors lead.} On the four datasets
with at least ten events, ASE, Hamming and NBS score higher than both versions of the counts. Their leads
over the causal counts, $0.09$ to $0.10$, depend on the test and on the scale; SCPD, BOCPD, TIRE and the SpotLight-style sketch are also above the causal counts, by $0.02$ to $0.04$, none significantly. Pooled over the four
datasets, only ASE's lead is significant, by the bootstrap, which decides there; it does not hold by the $t$
test or with re-dated stablecoin labels (\cref{app:inference}). With each scale taken from all earlier
snapshots of the segment, the \emph{expanding scale}, the three leads fall to $0.01$ to $0.03$. ASE and Hamming
also see membership (\cref{tab:sees}). Against vertex turnover, which sees only activity, the intervals of
both leads cover zero, so by a reading rule fixed before turnover was scored the leads are not separable
from turnover. Whether they see
structure can only be tested with changes of structure at fixed activity.

\subsection{Structural changes at fixed degree}
\label{sec:structure}

\paragraph{Planted changes.} To change structure at fixed activity, we replace $15$ consecutive snapshots
of a real series, from an onset on, by planted copies; a \emph{plant} changes each copy and keeps $n_+$, $|E|$
and the degree sequence; the rewirings also keep the degree of every vertex. Three plants rewire by double-edge swaps, aiming to move half the
edges: at random, towards pairs of similar degree (assortative), or so as to close triangles. The fourth
plant, \emph{relabelling}, exchanges the labels of disjoint pairs covering half the active vertices; it
changes only the labelled structure, so no permutation-invariant score can respond to it. A \emph{trial} is
one onset and one random seed. A method's \emph{shift} is the mean increase of its highest score over the
first five planted steps, compared with the same steps of the original series, in standard deviations of
its highest scores over event-free windows; following the analysis specification, a shift below $0.1$
standard deviations is negligible. Intervals need long series, so the results below come from HT09
and PAMAP2. On HT09 the triangle-closing plant moves almost no edge, which leaves five rewirings, three on
PAMAP2 and two on HT09; \cref{app:plants} gives the school network and MIT.

\begin{figure}[t]
  \centering
  \includegraphics[width=0.96\textwidth]{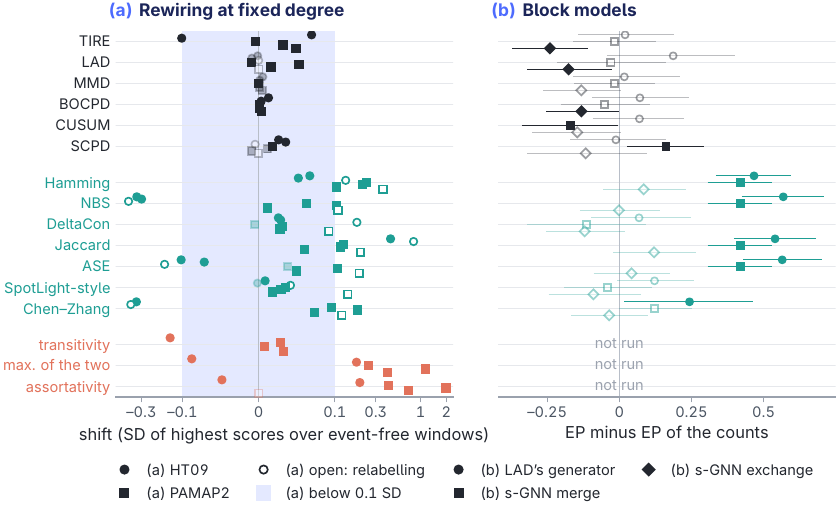}
  \caption{Structural changes at fixed degree. (a) Shift of each method's highest score after a plant on
  HT09 (circles) or PAMAP2 (squares), in standard deviations of its highest scores over event-free windows;
  filled: a rewiring; open: relabelling; faint: the $95\%$ interval of the paired AUC, the share of trials
  in which the planted score exceeds the original score, covers one half. The axis is linear
  within $\pm 0.1$ (shaded) and logarithmic beyond. Left out: plants that move fewer than $1\%$ of the edges,
  methods whose highest score is unchanged in at least half the trials (among them the counts), and TIRE on
  the relabelling, where it was not run. (b) EP minus the counts'
  EP, with $95\%$ intervals, in the block models of LAD and s-GNN at matched expected degree and in s-GNN's
  vertex exchange; filled: the interval excludes zero. Colours as in \cref{fig:main}.}
  \label{fig:structure}
\end{figure}

\paragraph{Permutation-invariant detectors barely respond.} Under the five rewirings, no
permutation-invariant detector shifts upward by $0.1$ standard deviations; the largest upward shift is
TIRE's, $0.07$ (\cref{fig:structure}(a)). Some of these small shifts are
consistent across trials but negligible. The only larger shift is downward: TIRE's score drops by $0.10$
under HT09's assortative rewiring
(\cref{app:plants}). Degree assortativity, a
permutation-invariant statistic built for one of these changes, shifts by $0.2$ to $2.0$ standard deviations
under four of the five. Relabelling leaves CUSUM, BOCPD, MMD, LAD and the structural statistics unchanged;
SCPD changes only through its random probe vectors, and TIRE was not run on it.

\paragraph{Several node-aligned detectors respond, partly to lost persistence.} Hamming shifts by more
than $0.2$ standard deviations under PAMAP2's random and assortative rewiring, and Jaccard by $0.45$ under
HT09's random rewiring; both also respond to relabelling. Each planted snapshot gets its own plant, so the
planted series loses \emph{edge persistence}, the tendency of the same pairs to be connected in
consecutive snapshots, and part of the response is to this loss. A persistent plant, applied once to all
fifteen snapshots, keeps persistence. With it the responses of NBS, Hamming, Jaccard and DeltaCon to HT09's
rewiring fall below $0.1$ standard deviations; on PAMAP2 the persistent rewiring moves almost no edge and
is not tested. Most responses to relabelling remain, among them Hamming's $0.50$ standard deviations
on PAMAP2 (\cref{app:plants}).

\paragraph{Block models at matched expected degree.} We hold the expected degree
fixed across each change in the generators of the LAD and s-GNN papers. In LAD's generator
($1{,}000$ vertices, mean degree $11$), whose changes alter the number of blocks, the counts have EP
$0.40$, not distinguishable from the constant score, and no permutation-invariant detector, LAD included,
differs significantly from the counts (EP $0.39$ to $0.59$), while NBS, Jaccard, Hamming and ASE reach EP $0.87$ to $0.97$ (\cref{fig:structure}(b)).
For local statistics this is expected: in sparse block models with equal expected degrees, the neighbourhood of a uniformly chosen vertex converges in distribution to the
Poisson Galton--Watson tree of the Erd\H{o}s--R\'enyi graph with that mean degree
\citep{vanderhofstad2024random}, in the sense of local weak convergence
\citep{benjamini2001recurrence,aldous2007processes}. Vertex averages of bounded local statistics therefore
have the same limit under all these models, and so do transitivity and degree assortativity under moment
conditions on the degrees. Extreme eigenvalues, which LAD uses and which are among the summaries of CUSUM, BOCPD, MMD and TIRE, fall outside this argument.
s-GNN's merge of four blocks into two
is denser (mean degree $110$) and outside the sparse limit; there SCPD, too, exceeds the counts by $0.16$,
as do four node-aligned detectors. In s-GNN's exchange of vertex pairs between blocks of equal density, a
change of labelled structure only, no method is significantly above the counts. Without matched degree both changes also alter the degree, and the
counts rank them at $0.97$ and $1.00$: such synthetic benchmarks reward a detector for seeing degree.

\section{Discussion and conclusion}
\label{sec:discussion}

\paragraph{Implications for evaluation.} On these benchmarks a detector's EP mixes its response to
activity with whatever it sees of structure. The counts respond only to the numbers of active vertices and
edges, a Rasch model motivates each of their terms as the score statistic for a change in one of these
numbers, and in an
exploratory comparison the events are large changes in them far more often than event-free windows. A
detector that claims to see structure therefore has to beat the counts, offline and causal; pooled over the
ten datasets none of the thirteen does so significantly, although on the four datasets with the most events
three node-aligned detectors are ahead of the counts, and seven detectors of the causal counts. Degree assortativity and several node-aligned detectors respond to rewiring that the counts cannot see,
the node-aligned detectors partly to lost persistence, while on the two long series the
permutation-invariant detectors barely respond. The real benchmarks lack labelled events that change
structure at fixed activity; until such events exist, tests at fixed degree are the direct evidence that a
detector sees structure.

\paragraph{Limitations.}\label{sec:limits} Six datasets have fewer than ten events, and the events of one
series share overlapping event-free windows but are resampled as independent, which can understate the
variance. The labels are noisy, the tolerances follow no common rule, six of the ten datasets were built or
added by us, and the settings of the counts and of five detectors were chosen on three datasets, MIT among them, which can favour settings that respond to activity. The node-aligned
detectors are two-window adaptations, binary graphs can handicap detectors built for weights or
directions, and we did not vary the correlation threshold of PAMAP2 and EEG. We ran no model-based detector
on the ten datasets, no graph-kernel test and no graph neural network (we could not reproduce s-GNN's results).

\begin{tcolorbox}[dsthm, top=7pt, bottom=7pt]
{\sffamily\bfseries\color{dsnavy}A protocol for a new detector}\par\vspace{3pt}
\begin{enumerate}[label=(\roman*),leftmargin=2.2em,itemsep=1pt,topsep=0pt,parsep=0pt]
  \item Report it beside the two counts, the causal counts and a constant score; report a node-aligned
  detector also beside vertex turnover.
  \item Rank each event among the event-free windows of its series and pool with equal weight per dataset,
  instead of reporting $F_1$ at a tuned threshold.
  \item Report the share of events whose activity change exceeds the $90$th percentile of event-free
  changes, beside the same share of event-free windows.
  \item Test at fixed degree, by rewiring with a relabelling control and persistent plants, and in block
  models at matched expected degree.
  \item Give a simultaneous bound on how far it could exceed the counts.
\end{enumerate}
\end{tcolorbox}

\bibliographystyle{plainnat}
\bibliography{references}

\appendix
\crefalias{section}{appendix}

\section{Detectors and settings}
\label{app:settings}

The
exploratory methods and analyses ($^\circ$) are turnover, fixed after the results on the four datasets with
at least ten events; the split by event type, tagged after the main results but before any EP by type was
computed, on datasets whose EP was known; and, chosen after their own results, the count variants, the equal
weight per event, the activity strata and the share of event-free windows that reach them, the common
tolerance, the label variants and shifts, the rules on event-free windows, the correlation-network subset
and the Enron series of \citet{peel2015detecting}. The reading rule for turnover (\cref{app:inference}) is
in the released code and is not time-stamped. At $\tau = 0$ a score that sees only $G_{t-1}$ and earlier
cannot score an event's first changed snapshot.

\begin{table}[!htb]
  \centering
  \caption{Scored methods (\cref{sec:protocol}): the setting of each and when it was fixed, \emph{before
  EP} (before any EP was computed), \emph{after first results} (after the counts' first results against
  the permutation-invariant detectors) or \emph{exploratory}. $F_1$: chosen by the highest mean $F_1$
  over MIT, Enron and AskUbuntu. $w$, $w_{\mathrm{ref}}$: test and reference windows; $w = 2$,
  $w_{\mathrm{ref}} = 10$ for the node-aligned detectors except SpotLight-style and for the structural
  statistics. $G_t$: the score at $t$ also sees $G_t$ (otherwise $G_{t-1}$ and earlier). Summaries, the
  eight inputs of TIRE, CUSUM, BOCPD and MMD: $n$, $|E|$, the mean and standard deviation of the degree
  over all $n$ vertices, the density, the number of connected components, and the largest absolute
  adjacency eigenvalue twice (as spectral radius and as largest singular value, equal for a symmetric
  matrix).}
  \label{tab:rows}
  \scriptsize
  \setlength{\tabcolsep}{3pt}
  \begin{tabular}{@{}lll@{}}
    \toprule
    Method & Setting & Fixed \\
    \midrule
    Counts & scale: whole series; $w = 2$, $w_{\mathrm{ref}} = 5$; $1/4$ on the squared edge term ($F_1$) & before EP \\
    Causal counts & scale: first $20$ steps of each segment & after first results \\
    Constant & the same value at every step & before EP \\
    \midrule
    CUSUM ($G_t$) & two-sided on the summaries; threshold $45$, $\kappa = 1.5$, start-up $10$, cooldown $7$ ($F_1$) & before EP \\
    BOCPD ($G_t$); MMD & on the summaries; hazard $0.1$; window $3$ ($F_1$) & before EP \\
    LAD ($G_t$); SCPD ($G_t$) & $5$ eigenvalues ($F_1$), windows $7$, $10$; $20$ moments, its paper's windows $5$, $10$ & before EP \\
    TIRE (offline) & on the summaries; window $5$, time and frequency domains, $100$ epochs ($F_1$) & before EP \\
    \midrule
    Hamming, Jaccard, NBS & $\ell_1$, weighted Jaccard, CUSUM at the split; window-mean adjacency matrices & after first results \\
    Chen--Zhang & edge-count statistic at the split on the Hamming minimum spanning tree & after first results \\
    DeltaCon; ASE & $g = 10$ vertex groups; distance of rank-$4$ spectral embedding estimates & after first results \\
    SpotLight-style & $k$-NN distance of the sketch of $G_{t-1}$ to the $10$ before; $50$ random subset pairs & after first results \\
    Structural statistics & scale: first $20$ steps of a segment & after first results \\
    \midrule
    Turnover & $\ell_1$ change of the 0/1 active-vertex vector, over $n$; $w = 2$, $w_{\mathrm{ref}} = 5$ & exploratory \\
    Variants & edge count alone, also causal; expanding scale; counts at $w_{\mathrm{ref}} = 10$; & exploratory \\
    & LAD with the $6$ or $499$ eigenvalues of its code; NetSimile & \\
    \bottomrule
  \end{tabular}
\end{table}

\paragraph{Grids and selection.} EP sees only the settings that change the score. Of the counts' $5{,}040$
settings only $48$ do ($16$ pairs of window lengths times three ways of combining the counts); the others
set thresholds and cooldowns, which EP ignores. The other tuned methods have $2{,}736$ such settings
(CUSUM), $360$ (TIRE), $60$ (LAD), $12$ (BOCPD) and $7$ (MMD). CUSUM's score resets after each alarm, so
its threshold changes the score; without resets its pooled EP is $0.563$, and the counts exceed it by
$+0.188$ [$0.054$, $0.336$]. Weights $1$, $1/2$ and $1/4$ on the squared edge term give pooled EP $0.751$,
$0.752$ and $0.751$; weight $0$ leaves the active-vertex count alone and gives $0.683$. The whole-series
scale is one constant per count and series, so it changes the ranking only through the counts' relative
weight. On the stablecoin network the first $20$ steps, which set the causal scale, contain one scored
event. At the windows of the node-aligned detectors ($w = 2$, $w_{\mathrm{ref}} = 10$) the counts have
pooled EP $0.738$ and stay above every detector. The selection used three MIT dates that differ from the
calendar dates of \cref{app:data}; reselecting with the calendar dates changes only the thresholds of CUSUM
and MMD and LAD's setting, and we keep the original selection.

\paragraph{Other rules for choosing settings.} Among the permutation-invariant methods the counts keep the
highest pooled EP under each of four other rules: the authors' defaults ($0.744$; the counts have no
published default and take the node-aligned detectors' windows), the median setting of each grid ($0.702$),
the best setting in hindsight ($0.762$) and, for each dataset, the setting best on the other nine ($0.751$;
Hamming and NBS, tuned by the same rule over their windows, have $0.655$ and $0.677$). On the four datasets
with at least ten events TIRE exceeds the counts when its setting is chosen in hindsight or on the other
datasets.

\paragraph{LAD's eigenvalues.} LAD keeps five Laplacian eigenvalues, chosen by $F_1$ from $5$
to $20$; its code uses $6$ on UCI and the Senate, $338$ on the Canadian bills and $499$ on its synthetic
data. With $6$ its pooled EP is $0.609$ (five: $0.610$). With $499$$^\circ$ it gains $0.085$ [$0.009$,
$0.161$] on the four datasets with at least ten events but falls on UCI ($0.91$ to $0.68$) and MAG ($0.79$
to $0.51$), and stays $0.136$ [$0.042$, $0.253$] below the counts on all ten (pooled EP $0.615$). The
signature distance of NetSimile$^\circ$ \citep{berlingerio2012netsimile} has $0.651$.

\paragraph{Not run.} We do not run MultiLAD \citep{huang2024laplacian}, the detectors of
\citet{ide2004eigenspace}, \citet{miller2020size} and \citet{wilson2019modeling}, embedding tests
\citep{zambon2019change}, AnomRank \citep{yoon2019fast}, the martingale detector of \citet{ho2026detecting}
or edge-stream detectors such as MIDAS \citep{bhatia2020midas}, surveyed by \citet{ranshous2015anomaly} and
\citet{akoglu2015graph}. GHRG runs only on its authors' series (\cref{app:data}).

\section{Data, events and per-dataset results}
\label{app:data}

\paragraph{Construction.} Every snapshot is binary and symmetric, without self-loops. Three datasets of
the source papers are not scored, because their events fall in the burn-in: the US Senate ($12$
Congresses, $225$ vertices, two events of collaboration extremes) and the Canadian bills ($14$ years,
$734$ vertices, two leadership changes) of LAD, and the COVID flights ($30$ weeks, $3{,}000$ vertices,
two travel bans) of SCPD; on the bills and the flights the events are the source detector's own
detections. UCI, the Senate and
the Canadian bills follow LAD's loaders \citep{opsahl2009clustering,huang2020laplacian}; MAG, the flights
and the stablecoin transfers follow SCPD's \citep{huang2023fast}. PAMAP2 \citep{reiss2012introducing}
follows s-GNN's: per subject, one correlation graph per $1$\,s epoch of the $27$ inertial channels, with an
edge where $|r| > 0.2$. The school and HT09 data \citep{stehle2011high,isella2011whats} are $5$-minute
contact graphs, one segment per day. EEG uses $2$\,s epochs of the bipolar channels of CHB-MIT cases
(patients) 1, 3 and 5 \citep{shoeb2009application}, with an edge where $|r| \ge 0.5$. AskUbuntu has one
answer graph per month, with Ubuntu releases as events \citep{miller2020size}; as the snapshots renumber
each month's users, the node-aligned detectors and turnover map them to $159{,}316$ user ids, and for
AskUbuntu the order of construction and scoring is not recorded.  The tolerances are ours except UCI's, zero, which is LAD's; the released
protocol holds those of the school, HT09, stablecoin, PAMAP2 and EEG. With $m$ event-free windows an
event's EP moves in steps of $1/(2m)$.

\paragraph{MIT and Enron.} MIT is the Social Evolution study \citep{madan2012sensing}: daily Bluetooth
proximity among $63$ students of one dormitory. We do not use the earlier Reality Mining study of the same
lab, used by \citet{peel2015detecting} and \citet{kendrick2018change}, because rebuilt it keeps only four
event-free windows after the burn-in. We keep the days with more than five records and more than one edge:
$270$ days from 30 January 2008 to 24 June 2009, all but eight of them from October 2008 on. From then on
the filter drops only 22 to 25 December and 3 January, so the Christmas event falls on 26 December. The six
events follow the MIT calendar for 2008--09 \citep{mit2008calendar} and the federal holidays: Columbus Day
(13 October 2008), Thanksgiving (27 November), Christmas, New Year, the spring vacation (23 March 2009) and
the day after the last day of classes (15 May). Over Christmas the activity drop starts one step before
the event. Enron has one weekly email graph on all addresses
\citep{klimt2004enron} and six corporate events merged into four.

\paragraph{Source labels.} SCPD's stablecoin labels are days of market news; we re-dated each against the
earliest dated public report. Of its $38$ labels (on $31$ days), $21$ move one day later, four two days
later and two one day earlier, five are unchanged and six are removed, which leaves $25$ scored events.
With these dates the counts stay significantly above BOCPD, MMD, SCPD and Chen--Zhang by the bootstrap,
and ASE's lead over the causal counts on the four datasets with at least ten events shrinks to $+0.083$
[$0.004$, $0.149$] (Holm-adjusted $p = 0.48$). On UCI the edges drop from $162$ to $12$ three snapshots
before the end-of-term label; we keep LAD's label, and \cref{app:inference} gives the pool without UCI.

\paragraph{Reproduction.} LAD's code and construction give Hits@10 (the share of labelled events among the
ten highest scores) $0.50$ on UCI and Hits@2 $1.00$ on the Senate, as reported. For s-GNN we used the authors' code (DyNNet, commit
2cd2035) with identity features, window $20$ and the default network, because the paper gives no settings
for PAMAP2. Its point-adjusted $F_1$, which credits a whole labelled interval for one alarm in it, has mean
$0.14$ over $16$ trainings. Tuned over the default and $12$ configurations of the paper's grid for its S\&P
500 data ($104$ trainings), it reaches $0.00$ at the configuration chosen by validation and $0.20$ at the
best in hindsight, against $0.78$ reported; the counts reach $0.84$, because point adjustment inflates every
score \citep{kim2022towards}.

\paragraph{Per-dataset results.} \Cref{tab:perds} gives every method's EP on each scored dataset; per
dataset, nominal $95\%$ intervals with two or four events covered only $50\%$ and $77\%$ of the time in a
simulation, so single-dataset statements on UCI, MAG, Enron and the school network carry no weight.
\Cref{tab:own} asks whether our binary construction handicaps the source detectors: on their own
constructions they lead the counts on three of the six series, the Canadian bills, MAG and the flights,
and only MAG has more than two events. On our Enron the counts ($0.777$) trail six methods and turnover, among them TIRE ($0.934$) and CUSUM
($0.875$).

\begin{table}[!t]
  \begin{minipage}[t]{0.6\textwidth}
  \centering\captionsetup{justification=raggedright,singlelinecheck=false}
  \caption{EP of every method on each scored dataset (scored events in parentheses); rows as in
  \cref{tab:pooled}. Bold: the highest detector or structural statistic per column (ties all bold). On
  PAMAP2 DeltaCon ($0.921$) is above the counts ($0.916$). Last two rows: share of large or extreme
  activity changes (\cref{sec:events}) among the events and among the event-free windows, each window
  ranked like an event.}
  \label{tab:perds}
  \scriptsize
  \setlength{\tabcolsep}{0.9pt}
  \begin{tabular}{lcccccccccc}
    \toprule
    & UCI & MAG & Stablec. & PAMAP2 & MIT & Enron & School & HT09 & EEG & AskU. \\
    & (2) & (5) & (30) & (97) & (6) & (4) & (4) & (22) & (19) & (9) \\
    \midrule
    Counts & .81 & .61 & .59 & .92 & .96 & .78 & .91 & .67 & .69 & .58 \\
    Causal counts & .83 & .68 & .59 & .79 & .95 & .78 & .88 & .52 & .68 & .58 \\
    \midrule
    TIRE & .30 & .63 & .46 & .82 & .87 & \textbf{.93} & .53 & .59 & .81 & .51 \\
    LAD & \textbf{.91} & .79 & .50 & .79 & .76 & .46 & .44 & .41 & .56 & .47 \\
    MMD & .88 & .41 & .43 & .71 & .78 & .49 & .70 & .53 & .68 & .43 \\
    BOCPD & .61 & .37 & .53 & .91 & .89 & .53 & .30 & .62 & .67 & .44 \\
    CUSUM & .30 & .51 & .61 & .61 & .28 & .88 & \textbf{.86} & .56 & .59 & .53 \\
    SCPD & .23 & .43 & .60 & .87 & .60 & .72 & .63 & .72 & .54 & .27 \\
    \midrule
    Hamming & .48 & .55 & .65 & .91 & .45 & .88 & .78 & .67 & .75 & .52 \\
    NBS & .43 & .46 & .55 & .88 & .64 & .88 & .56 & .65 & \textbf{.84} & .67 \\
    DeltaCon & .68 & .72 & .23 & \textbf{.92} & \textbf{.90} & .91 & .40 & .69 & .60 & .48 \\
    Jaccard & .56 & .79 & .46 & .76 & .81 & .32 & .78 & .52 & .70 & .56 \\
    ASE & .22 & .68 & \textbf{.68} & .91 & .47 & .68 & .69 & .66 & .74 & .41 \\
    SpotLight-style & .20 & .65 & .54 & .74 & .76 & .31 & .83 & \textbf{.75} & .62 & \textbf{.68} \\
    Chen--Zhang & .13 & .46 & .44 & .77 & .34 & .37 & .61 & .52 & .82 & .64 \\
    \midrule
    Transitivity & .50 & .74 & .52 & .82 & .77 & .82 & .73 & .49 & .70 & .50 \\
    Max.\ of the two & .45 & \textbf{.89} & .42 & .76 & .85 & .35 & .77 & .48 & .51 & .49 \\
    Assortativity & .46 & \textbf{.89} & .44 & .61 & .88 & .40 & .83 & .43 & .47 & .49 \\
    \midrule
    Turnover$^\circ$ & .57 & .44 & .61 & .75 & .88 & .87 & .63 & .72 & .63 & .54 \\
    \midrule
    Events$^\circ$ & .50 & .20 & .27 & .78 & .67 & .50 & .50 & .32 & .42 & .33 \\
    Event-free$^\circ$ & .14 & .20 & .15 & .10 & .15 & .14 & .18 & .12 & .18 & .10 \\
    \bottomrule
  \end{tabular}
  \end{minipage}\hfill
  \begin{minipage}[t]{0.38\textwidth}
  \centering\captionsetup{justification=raggedright,singlelinecheck=false}
  \caption{Each source detector (LAD's datasets, then SCPD's) and the counts on the detector's own
  construction (Own) and on our binary graphs (Binary), on the events both score (from step $7$ on the
  Senate, bills and flights), so values can differ from \cref{tab:perds}. Rows with one or two events
  are descriptive.}
  \label{tab:own}
  \scriptsize
  \setlength{\tabcolsep}{2.6pt}
  \begin{tabular}{llccccc}
    \toprule
    & & & \multicolumn{2}{c}{Own} & \multicolumn{2}{c}{Binary} \\
    \cmidrule(lr){4-5}\cmidrule(l){6-7}
    Dataset & Det. & Ev. & Det. & Counts & Det. & Counts \\
    \midrule
    UCI & LAD & 2 & .497 & .792 & .908 & .808 \\
    Senate & LAD & 1 & 1.00 & 1.00 & .500 & 1.00 \\
    Bills & LAD & 2 & 1.00 & .500 & .300 & .500 \\
    MAG & SCPD & 5 & .850 & .625 & .452 & .625 \\
    Flights & SCPD & 1 & 1.00 & .063 & 1.00 & .063 \\
    Stablecoin & SCPD & 30 & .394 & .590 & .601 & .590 \\
    \bottomrule
  \end{tabular}
  \end{minipage}
\end{table}

\paragraph{Peel and Clauset's series.} GHRG is the one model-based detector we run, on the two series of
\citet{peel2015detecting}, which we rebuilt: MIT Reality Mining \citep{eagle2006reality} and Enron with
$151$ addresses and $23$ events (log weekly edge counts of our Enron and theirs correlate at $0.925$). On
their Enron at a tolerance of one week the counts have EP $0.563$; CUSUM ($0.710$), SCPD ($0.705$), TIRE
($0.610$), GHRG's likelihood ratio ($0.595$) and LAD ($0.576$) are above them, and BOCPD, GHRG's
significance level, MMD and the authors' baselines ($0.306$ to $0.481$) below. Without GHRG, across both Enron
constructions, event lists and tolerances of one and four weeks, CUSUM leads in seven of eight cases and
the counts rank second to fourth of seven methods. Their Reality Mining series keeps four event-free
windows, too few to report alone. Added to the ten datasets, the two series give the counts a pooled EP of
$0.713$ (TIRE next: $0.658$).

\section{Proof of the score identity}
\label{app:proofs}

\paragraph{The identity.} Each present unit of a snapshot adds one to $\sum_i c_i$, and one to
$\sum_i x_i$ if the snapshot is in the test window, so $\sum_i x_i = w\,\bar n_{\mathrm{test}}$ and
$\sum_i c_i = w\,\bar n_{\mathrm{test}} + w_{\mathrm{ref}}\,\bar n_{\mathrm{ref}}$. For any 0/1
presence array, therefore, $\sum_i (x_i - w c_i/N) = w\,\bar n_{\mathrm{test}}(1 - w/N) -
(w\,w_{\mathrm{ref}}/N)\,\bar n_{\mathrm{ref}} = (w\,w_{\mathrm{ref}}/N)(\bar n_{\mathrm{test}} -
\bar n_{\mathrm{ref}})$, because $1 - w/N = w_{\mathrm{ref}}/N$.

\paragraph{The score statistic.} Under the Rasch model, fix a unit $i$ and condition on $c_i$. Each
set of $c_i$ snapshots that could contain $i$ then has probability proportional to $e^{\beta x_i}$,
where $x_i$ counts the test snapshots in the set, and $\alpha_i$ cancels. The conditional likelihood
is $\prod_i \binom{w}{x_i}\binom{w_{\mathrm{ref}}}{c_i - x_i} e^{\beta x_i}/Z_i(\beta)$, where
$Z_i(\beta)$ sums the numerator over $x_i$. At $\beta = 0$, $x_i$ is hypergeometric with mean
$\partial_\beta \log Z_i(0) = w c_i/N$, so the score at $\beta = 0$ is $\sum_i (x_i - w c_i/N)$,
the left side of \eqref{eq:score}.

\paragraph{The standardized test.} The null law of $x_i$ given $c_i$ assumes that the snapshots
are exchangeable for unit $i$; edge persistence or a trend in activity breaks this. The product likelihood
above and the conditional variance of the score, the sum of the hypergeometric variances
$\sum_i c_i (N - c_i)\,w\,w_{\mathrm{ref}}/(N^2(N-1))$, assume independent units; pairs that share a vertex
need not be independent, and the score then remains the score of the independence likelihood, but this
variance no longer holds.
Standardized by this variance, the score test has pooled EP $0.751$ with pairs as units, the same as
the counts (counts minus test $0.000$ [$-0.051$, $0.048$]), $0.716$ with vertices as units and
$0.679$ for the joint test, the two standardized scores weighted by their inverse covariance.

\section{Inference and robustness}
\label{app:inference}

\paragraph{Coverage and power.} In populations of ten datasets drawn from the observed ones, the two-stage
interval of counts minus TIRE covered in $91\%$ of the replicates, a $t$ interval in $94\%$, and an
interval that resamples events only in $51\%$. With four datasets the two-stage and $t$ intervals covered
in $96\%$, an optimistic estimate; the released specification keeps the bootstrap's Holm marks on the four
datasets when this coverage is at least $90\%$, and reports the $t$ test beside them. An unadjusted test
detects a lead with $80\%$ power when the lead is $2.80$ two-stage standard deviations; depending on the
method, this is $0.12$ to $0.25$ in pooled EP on the ten datasets and $0.09$ to $0.23$ on the four.

\paragraph{Tests and bounds.} Every method is scored on the same $10{,}000$ bootstrap resamples, so the
differences are paired. The counts' pooled EP has the interval [$0.655$, $0.845$] and the causal counts'
[$0.633$, $0.822$]. The sign-flip test is exact if the per-dataset differences are symmetric about zero;
with ten datasets its smallest $p$ is $0.002$, or $0.031$ after Holm's correction over sixteen methods. By
the bootstrap, BOCPD, MMD and SCPD are below the counts after Holm's correction; by the exact test only
BOCPD is (MMD $0.088$, SCPD $0.19$). Against the causal counts only MMD is below, by the bootstrap
(Holm-adjusted $p = 0.048$). BOCPD and MMD read $|E|$ among their summaries, so two of these contrasts
compare two uses of the same count. The counts stay first when any one dataset is left out (pooled EP $0.728$ to $0.770$), as
do the causal counts (from $0.704$). By the exact sign-flip test with Bonferroni's correction, no method's
lead over the counts exceeds $0.128$; by a $t$ bound, $0.130$; by the bootstrap's maximum studentized
difference, $0.072$. Over the causal counts the three bounds are $0.145$, $0.177$ and $0.104$;
\cref{sec:events} quotes the larger dataset-level bound of each. On the four datasets with at least ten
events the exact bound is infinite, the Bonferroni $t$ bound $0.505$ and the bootstrap bound $0.137$.

\begin{table}[!htb]
  \centering
  \caption{Pooled EP beyond \cref{tab:pooled}: the four datasets with at least ten events and equal weight per
  event, with paired two-stage $95\%$ intervals. $\ge 10$ events (HT09,
  stablecoin, PAMAP2, EEG): EP and differences from the counts and the causal counts (negative: the
  method is above); $\dagger$: significant after Holm's correction by the two-stage bootstrap, whose
  marks the specification keeps on these datasets; by the $t$ test only the larger of the two structural
  changes (counts). Equal weight per event (all ten; PAMAP2 then has
  half the weight). Rows as in \cref{tab:pooled}, then the variants ($^\circ$: as in \cref{sec:protocol}), whose pooled
  EP with equal weight per dataset is in the Variants paragraph and \cref{app:settings}. Bold: the
  highest detector or structural statistic per column. Expanding: scaled by all earlier steps of the segment. LAD,
  $499$ eigenvalues: LAD with the number its code uses.}
  \label{tab:pooledfull}
  \scriptsize
  \setlength{\tabcolsep}{3pt}
  \begin{tabular}{lcccc}
    \toprule
    & \multicolumn{3}{c}{$\ge 10$ events} & Equal weight per event \\
    \cmidrule(lr){2-4}\cmidrule(l){5-5}
    Method & EP & Counts $-$ method & Causal counts $-$ method & EP, all ten \\
    \midrule
    Counts & 0.716 & -- & $-0.071$ [$-0.146$, 0.000] & 0.791 \\
    Causal counts & 0.645 & $+0.071$ [$-0.000$, 0.146] & -- & 0.714 \\
    Constant & 0.500 & $+0.216$ [0.099, 0.354] & $+0.145$ [0.033, 0.257] & 0.500 \\
    \midrule
    \multicolumn{5}{l}{\emph{Permutation-invariant detectors (TIRE, CUSUM, BOCPD, MMD on snapshot summaries)}} \\
    TIRE (offline) & 0.671 & $+0.045$ [$-0.082$, 0.149] & $-0.026$ [$-0.133$, 0.094] & 0.714 \\
    LAD & 0.566 & $+0.150$ [0.068, 0.250]$^\dagger$ & $+0.079$ [0.002, 0.166] & 0.655 \\
    MMD & 0.587 & $+0.129$ [0.033, 0.208] & $+0.058$ [$-0.031$, 0.145] & 0.623 \\
    BOCPD & 0.682 & $+0.034$ [$-0.027$, 0.101] & $-0.037$ [$-0.140$, 0.062] & 0.737 \\
    CUSUM & 0.592 & $+0.123$ [0.003, 0.256] & $+0.052$ [$-0.042$, 0.152] & 0.592 \\
    SCPD & 0.683 & $+0.032$ [$-0.060$, 0.126] & $-0.039$ [$-0.168$, 0.094] & 0.721 \\
    \midrule
    \multicolumn{5}{l}{\emph{Node-aligned detectors (two-window adaptations)}} \\
    Hamming & 0.743 & $-0.027$ [$-0.095$, 0.031] & $-0.098$ [$-0.161$, $-0.033$] & \textbf{0.778} \\
    NBS & 0.730 & $-0.015$ [$-0.115$, 0.064] & $-0.086$ [$-0.168$, 0.010] & 0.764 \\
    DeltaCon & 0.608 & $+0.108$ [$-0.023$, 0.289] & $+0.037$ [$-0.154$, 0.267] & 0.719 \\
    Jaccard & 0.611 & $+0.105$ [0.009, 0.184] & $+0.034$ [$-0.049$, 0.122] & 0.666 \\
    ASE & \textbf{0.749} & $-0.033$ [$-0.102$, 0.026] & $-0.104$ [$-0.157$, $-0.045$]$^\dagger$ & 0.774 \\
    SpotLight-style & 0.664 & $+0.052$ [$-0.052$, 0.154] & $-0.019$ [$-0.165$, 0.096] & 0.684 \\
    Chen--Zhang & 0.638 & $+0.078$ [$-0.071$, 0.195] & $+0.007$ [$-0.115$, 0.128] & 0.654 \\
    \midrule
    \multicolumn{5}{l}{\emph{Structural statistics}} \\
    Transitivity & 0.635 & $+0.080$ [$-0.004$, 0.169] & $+0.009$ [$-0.052$, 0.078] & 0.706 \\
    Max.\ of the two & 0.540 & $+0.176$ [0.114, 0.247]$^\dagger$ & $+0.105$ [0.021, 0.197] & 0.633 \\
    Assortativity & 0.488 & $+0.228$ [0.138, 0.310]$^\dagger$ & $+0.157$ [0.083, 0.240]$^\dagger$ & 0.560 \\
    \midrule
    Turnover$^\circ$ & 0.679 & $+0.037$ [$-0.052$, 0.132] & $-0.034$ [$-0.154$, 0.065] & 0.700 \\
    Causal counts, expanding$^\circ$ & 0.719 & $-0.003$ [$-0.025$, 0.022] & $-0.074$ [$-0.146$, $-0.004$] & 0.800 \\
    Edge count$^\circ$ & 0.710 & $+0.005$ [$-0.032$, 0.042] & $-0.066$ [$-0.132$, $-0.000$] & 0.790 \\
    Counts, $w_{\mathrm{ref}} = 10^\circ$ & 0.716 & $+0.000$ [$-0.041$, 0.038] & $-0.071$ [$-0.163$, 0.016] & 0.788 \\
    Causal edge count$^\circ$ & 0.686 & $+0.029$ [$-0.012$, 0.074] & $-0.041$ [$-0.097$, 0.024] & 0.764 \\
    LAD, $499$ eigenvalues$^\circ$ & 0.650 & $+0.065$ [$-0.002$, 0.148] & $-0.006$ [$-0.108$, 0.095] & 0.736 \\
    \bottomrule
  \end{tabular}
\end{table}

\paragraph{The four datasets with at least ten events.} On HT09, the stablecoin network, PAMAP2 and EEG,
ASE, Hamming and NBS are above the counts with Holm-adjusted $p = 1.00$ by the bootstrap and by $t$ tests
(\cref{tab:pooledfull}). ASE's lead over the causal counts is $+0.104$: by the bootstrap its interval is
[$0.045$, $0.157$] and its Holm-adjusted $p$ $0.03$; by the $t$ test its interval is [$0.051$, $0.156$] and
its Holm-adjusted $p$ $0.126$. Hamming's lead is $+0.098$ (Holm-adjusted $p$ $0.076$ by the bootstrap,
$0.259$ by $t$); no sign-flip test on four datasets can give $p$ below $0.125$. ASE's result is fragile: with the audited stablecoin dates its Holm-adjusted $p$ is $0.48$,
and with a stage that resamples PAMAP2's eight subjects and EEG's three patients $0.46$; on the ten
datasets that stage barely moves the intervals. Leaving one dataset out moves the significance between detectors: without the
stablecoin network NBS's lead over the causal counts, $+0.126$ [$0.070$, $0.195$], is significant after
Holm's correction and ASE's and Hamming's are not (Holm-adjusted $p = 0.107$ each); without EEG, ASE's ($+0.117$ [$0.063$,
$0.168$]) and Hamming's ($+0.107$, $p = 0.045$) are, and NBS's is not. The causal scale is the reason: the first $20$ steps of a
segment are a poor reference on HT09, where each day starts with the morning arrivals, and on PAMAP2.
Against the counts scaled by all earlier steps of the segment (the expanding scale$^\circ$; pooled EP
$0.757$ on the ten and $0.719$ on the four), ASE's and Hamming's leads fall to $+0.030$ [$-0.047$, $0.107$]
and $+0.024$ [$-0.038$, $0.086$] ($t$ intervals).

\paragraph{Vertex turnover.} ASE and Hamming see membership as well as structure (\cref{tab:sees}), so after
their leads on these four datasets we scored turnover, which sees membership but no structure. A reading
rule written before it was scored says that their leads are not separable from turnover if the intervals of
both minus turnover cover zero. Turnover has $0.679$ there, and ASE and Hamming exceed it by $+0.070$
[$-0.028$, $0.165$] and $+0.064$ [$-0.029$, $0.163$], so by the rule their leads are not separable from
turnover, though with four datasets the rule has little power. On PAMAP2 alone they exceed it by about
$0.16$, with intervals above zero; on HT09 turnover is above ASE by $0.061$ [$0.008$, $0.115$] (event-level
interval).

\paragraph{Event types.} An event is activity by design if any of its descriptions contains break, holiday,
vacation, Christmas, New Year, Thanksgiving, Columbus, lunch, coffee, term, classes, session, workshop,
keynote, volume (or vol) or ``whales leave''; the three stablecoin events so tagged report record trading
volumes or large holders leaving, so the stablecoin network has events of both types. With five and six
datasets the exact test cannot resolve the split (smallest unadjusted $p$ $0.0625$ and $0.031$). By the
bootstrap with Holm's correction the counts are significantly above eleven of the sixteen methods on the
activity-by-design events and the causal counts above three; on the other events no method is
significantly above either version, and the counts have the highest EP in $4\%$ of the resamples. Counted
as activity by design, AskUbuntu's releases let Hamming lead the counts on the other events ($0.745$
against $0.712$). After the split was seen$^\circ$: MIT, UCI and the school network add $0.097$ to the
counts' pooled lead over Hamming, which is $0.087$ in all; on HT09, whose events are all activity by design,
the counts do not lead; and the tag does not track the measured activity change ($16$ of the $37$
activity-by-design events and $96$ of the $161$ others are large or extreme, $76$ of the latter on PAMAP2).
\Cref{tab:pools} gives the counts on these and other subsets.

\begin{table}[!htb]
  \centering
  \caption{The counts on subsets of the events (upper block) and of the datasets (lower block),
  equal weight per dataset. Highest other method: the highest detector or structural statistic, chosen
  after the fact; counts minus that method with a two-stage $95\%$ interval. Each block by descending
  counts minus method. Other variants: \cref{tab:robust}.}
  \label{tab:pools}
  \footnotesize
  \setlength{\tabcolsep}{4pt}
  \begin{tabular}{lcclc}
    \toprule
    Datasets or events (datasets; events) & Counts & Causal counts & Highest other method & Counts $-$ method \\
    \midrule
    Activity-by-design events ($5$; $37$) & 0.827 & 0.795 & MMD 0.695 & $+0.133$ [$-0.008$, 0.302] \\
    Other events ($6$; $161$) & 0.690 & 0.680 & NBS 0.713 & $-0.024$ [$-0.122$, 0.074] \\
    Held out, other events ($4$; $148$) & 0.696 & 0.680 & ASE 0.752 & $-0.056$ [$-0.127$, 0.008] \\
    Source datasets, other events ($3$; $129$) & 0.699 & 0.681 & ASE 0.756 & $-0.057$ [$-0.135$, 0.009] \\
    \midrule
    Held out from the settings ($7$; $179$) & 0.742 & 0.711 & Hamming 0.684 & $+0.058$ [$-0.033$, 0.170] \\
    Built or added by us ($6$; $64$) & 0.763 & 0.731 & TIRE 0.707 & $+0.055$ [$-0.100$, 0.224] \\
    Source datasets ($4$; $134$) & 0.733 & 0.725 & LAD 0.750 & $-0.017$ [$-0.190$, 0.121] \\
    Correlation networks ($2$; $116$) & 0.801 & 0.733 & NBS 0.861 & $-0.060$ [$-0.209$, 0.042] \\
    \bottomrule
  \end{tabular}
\end{table}

\paragraph{Activity strata.} An event's activity change is ranked among the event-free windows of its
series, none of which overlaps it; each event-free window is ranked the same way, among the event-free
windows that do not overlap it. For one count the share of event-free windows above the $90$th percentile is
then close to $10\%$ by construction ($10\%$ for $n_+$, $11\%$ for $|E|$, averaged over datasets), and for
either count $15\%$ (last rows of \cref{tab:perds}). The share of events is higher on nine datasets and
slightly lower on MAG ($0.200$ against $0.201$), exceeds one half only on MIT and PAMAP2, and is $41\%$ on the eight datasets that are not
correlation networks. Pooled over events it is $112$ of $198$, $76$ of them PAMAP2's. On the $86$ ordinary
events the counts ($0.623$) and Hamming ($0.594$) do not differ significantly. Within each of the seven
datasets with both ordinary and extreme events, EP on the extreme minus the ordinary events$^\circ$ is
positive on average for every method except SCPD ($-0.043$); five of these datasets have one to three
extreme events, so this contrast is descriptive.

\paragraph{Variants.} \Cref{tab:robust} repeats the pooled comparison under each doubt about labels,
event-free windows and alignment. With one tolerance on every dataset$^\circ$ (without AskUbuntu from $\tau = 3$ on, where it keeps no event-free window) both versions of the counts
stay above every detector and structural statistic from $\tau = 1$ to $10$ (counts $0.67$ to $0.79$),
turnover is above the causal counts at $\tau = 9$ and $10$, and TIRE leads at $\tau = 0$ ($0.62$ against
$0.59$); on the four datasets with at least ten events another method leads at every tolerance. Of the label
shifts from two steps earlier to two later, the only one where another method leads is the shift two steps
earlier. With equal weight per event (\cref{tab:pooledfull}) the counts have $0.791$, above Hamming
($0.778$), ASE ($0.774$), NBS ($0.764$), BOCPD ($0.737$), SCPD ($0.721$) and DeltaCon ($0.719$), while the
causal counts ($0.714$) fall below these six and level with TIRE; none of the six is significantly above
them by the two-stage bootstrap. On the two correlation networks$^\circ$ NBS is above the causal counts by
$0.127$ [$0.070$, $0.217$], but with two datasets the dataset stage has almost nothing to resample. The
count variants of \cref{tab:pooledfull} are within $0.023$ of the counts on all ten datasets (expanding
scale $0.757$, edge count $0.742$ [$0.662$, $0.820$], $w_{\mathrm{ref}} = 10$ $0.738$, causal edge count
$0.728$), so $|E|$ carries most of the result.

\begin{table}[!htb]
  \centering
  \caption{The pooled comparison under variants beyond \cref{tab:pools}, each named by the concern it
  addresses: the counts, the highest detector or structural statistic and counts minus it (two-stage
  $95\%$ interval). Ten datasets unless a row names nine or the four with at least ten events. Upper
  block: the counts lead; lower block: another method leads; every interval covers $0$. The label
  variants and the rows on event-free windows are exploratory.}
  \label{tab:robust}
  \scriptsize
  \setlength{\tabcolsep}{4pt}
  \begin{tabular}{llll}
    \toprule
    Concern: variant & Counts & Highest other method & Counts $-$ method \\
    \midrule
    Labels from detections: without MAG & 0.766 & NBS 0.678 & $+0.088$ [$-0.044$, 0.234] \\
    Label after the drop: without UCI & 0.744 & Hamming 0.685 & $+0.059$ [$-0.046$, 0.199] \\
    Stablecoin dates: audited & 0.757 & Hamming 0.668 & $+0.089$ [$-0.023$, 0.225] \\
    Stablecoin dates: without the stablecoin network & 0.769 & DeltaCon 0.699 & $+0.070$ [$-0.046$, 0.214] \\
    Mixed scales: event-free windows of the event's subject or case & 0.752 & Transitivity 0.676 & $+0.076$ [$-0.008$, 0.160] \\
    Event-free windows: none $\le 17$ steps after an event (nine datasets) & 0.780 & DeltaCon 0.677 & $+0.103$ [$-0.042$, 0.270] \\
    \quad none $\le w + w_{\mathrm{ref}}$ of each method after an event (nine) & 0.779 & DeltaCon 0.678 & $+0.101$ [$-0.041$, 0.270] \\
    Alignment: every score at $t$ uses $G_t$ & 0.787 & NBS 0.698 & $+0.089$ [$-0.018$, 0.197] \\
    \midrule
    $\ge 10$ events, audited stablecoin dates & 0.731 & Hamming 0.752 & $-0.021$ [$-0.090$, 0.037] \\
    $\ge 10$ events without the stablecoin network & 0.757 & NBS 0.790 & $-0.033$ [$-0.150$, 0.055] \\
    Label dates: every event two steps earlier & 0.634 & TIRE 0.657 & $-0.023$ [$-0.124$, 0.067] \\
    Tolerance: $\tau = 0$ & 0.592 & TIRE 0.624 & $-0.032$ [$-0.153$, 0.121] \\
    \bottomrule
  \end{tabular}
\end{table}

\paragraph{Residual design.} An exploratory design
that regressed each score on the window changes of $n_+$ and $|E|$ and ranked the residuals failed its
calibration: the residuals of the constant score and of the counts, which should rank events at chance, have
EP $0.589$ [$0.543$, $0.632$] and $0.615$, so we report no residual EP.

\section{Planted and simulated structural changes}
\label{app:plants}

\paragraph{Plants.} Every rewiring is a sequence of double-edge swaps, $(a,b),(c,d) \to (a,d),(c,b)$ on
four distinct vertices whose new edges are absent; a swap keeps every degree. A plant accepts swaps until
$\rho|E|/2$ are accepted, with $\rho = 0.5$, or stops after $50$ times as many tries, so the rewirings move
$14\%$ to $40\%$ of the edges and HT09's triangle-closing plant almost none. The random plant accepts every
valid swap; the assortative plant only swaps with $d_ad_d + d_cd_b > d_ad_b + d_cd_d$, which raise degree
assortativity at fixed degrees; the triangle-closing plant only swaps that add triangles. The relabelling
draws, in snapshot $t$, $\mathrm{round}(\rho\, n_+(t)/2)$ disjoint pairs of active vertices and exchanges
the labels of each pair. The persistent variant draws one rewiring, or one relabelling of the vertices active
at the onset, and applies it to all $15$ snapshots. Onsets are $40$ steps apart, with offsets $0$ to $39$ and
ten seeds; a \emph{cell} is one series with one plant. The \emph{paired AUC} of a cell is the share of trials
in which a method's highest planted score exceeds its highest original score at the same steps, ties counted
as one half, with a $95\%$ interval from a bootstrap over $40$-step periods. A method tied in at least half
the trials of a cell (the counts: all) is not evaluable there; cells with fewer than ten periods, those of
the school network (four) and MIT (six), are descriptive, and HT09 has ten and PAMAP2 $478$. TIRE was run only on seeds $0$ to $2$ and
offsets $0$ and $20$, and not on the relabelling.

\paragraph{Shift sizes.} The released specification calls a response consistent but negligible when its
paired interval lies above one half, its unpaired AUC is below $0.55$ and its mean shift is below $0.1$
standard deviations; \cref{sec:structure} uses the last condition. In the five rewiring cells of HT09 and
PAMAP2 that move edges, TIRE, CUSUM (counted by the released rule, although tied in at least half the trials) and SCPD shift consistently in three, BOCPD and LAD in two and MMD in one,
each by less than $0.1$ standard deviations; the only larger shift is TIRE's downward shift of $0.102$ under HT09's assortative
rewiring. Degree assortativity shifts by $0.196$ under HT09's assortative rewiring and by $0.418$, $1.976$
and $0.722$ under PAMAP2's three rewirings, and falls under HT09's random rewiring. Under HT09's random
rewiring Jaccard shifts by $0.446$, Hamming's upward shifts are negligible, and NBS and Chen--Zhang fall by
about a third of a standard deviation. On PAMAP2 Hamming shifts by $0.233$ and $0.210$ under
random and assortative rewiring. On the school network, too short for intervals, SCPD and TIRE shift by
$0.39$ and $0.56$ under random rewiring, where the larger of the two structural changes shifts by $3.08$.

\paragraph{Relabelling controls.} On the relabelling cells the counts, CUSUM, BOCPD, MMD, LAD and the
structural statistics give the planted and the original series the same score in every trial, to a relative
difference below $10^{-9}$; LAD's school interval [$0.352$, $0.479$] excludes $0.5$ only through this
round-off. SCPD's school interval excludes $0.5$ because SCPD draws its random probe vectors in vertex order,
so a relabelling changes its draws; with probes redrawn on unchanged series, $4$ of $40$ intervals exclude
$0.5$.

\begin{table}[!htb]
  \centering
  \caption{Paired AUC of node-aligned detectors, plants per snapshot (independent, the design of
  \cref{sec:structure})
  / once for all $15$ snapshots (persistent), $\rho = 0.5$; brackets: persistent $95\%$ interval.
  Cells: series and plant (R random, A assortative, T triangle-closing, S relabelling).
  Moved: median over trials of the mean share of the original edges that the plant removes over the $15$
  planted snapshots, in the plants regenerated for the node-aligned detectors (the plants of
  \cref{fig:structure}(a) differ on HT09). School: four
  periods, descriptive. Grey: not evaluable (tied in at least half the trials).}
  \label{tab:persist}
  \scriptsize
  \setlength{\tabcolsep}{3pt}
  \begin{tabular}{llllll}
    \toprule
    Cell & Moved & Hamming & NBS & ASE & Chen--Zhang \\
    \midrule
    School R & .39/.22 & .87 / .80 [.79, .83] & .19 / .79 [.78, .81] & .18 / .20 [.18, .24] & .25 / \textcolor{dsgrey}{.58 [.46, .67]} \\
    School A & .39/.20 & .86 / .78 [.76, .82] & .19 / .75 [.73, .75] & .21 / .22 [.18, .26] & .27 / \textcolor{dsgrey}{.58 [.45, .67]} \\
    School T & .14/.07 & .71 / .44 [.40, .52] & .28 / .69 [.67, .72] & .76 / .79 [.73, .82] & \textcolor{dsgrey}{.46} / \textcolor{dsgrey}{.56 [.51, .61]} \\
    School S & .72/.68 & .95 / .95 [.93, .96] & .23 / .98 [.93, 1.00] & .20 / .81 [.75, .89] & .20 / .58 [.44, .72] \\
    HT09 R & .40/.18 & .79 / \textcolor{dsgrey}{.64 [.61, .67]} & .21 / .56 [.54, .58] & .28 / .44 [.41, .48] & .31 / \textcolor{dsgrey}{.46 [.44, .48]} \\
    HT09 A & .19/.13 & .72 / \textcolor{dsgrey}{.60 [.56, .65]} & .27 / \textcolor{dsgrey}{.57 [.54, .60]} & .37 / .55 [.49, .60] & \textcolor{dsgrey}{.38} / \textcolor{dsgrey}{.47 [.45, .49]} \\
    HT09 T & .05/.03 & \textcolor{dsgrey}{.64} / \textcolor{dsgrey}{.52 [.49, .56]} & \textcolor{dsgrey}{.39} / \textcolor{dsgrey}{.54 [.51, .56]} & \textcolor{dsgrey}{.49} / \textcolor{dsgrey}{.58 [.52, .62]} & \textcolor{dsgrey}{.47} / \textcolor{dsgrey}{.50 [.49, .50]} \\
    HT09 S & .68/.47 & .83 / .79 [.76, .82] & .20 / .76 [.68, .82] & .26 / .54 [.49, .58] & .28 / \textcolor{dsgrey}{.47 [.43, .51]} \\
    \bottomrule
  \end{tabular}
\end{table}

\paragraph{Persistence.} With one plant, NBS rises above $0.5$ in the school and HT09 cells that are not tied,
Chen--Zhang moves to about $0.5$, and Hamming stays above $0.5$ on both relabellings, so its response is not
only to lost persistence. Under HT09's persistent relabelling Hamming's shift falls to $0.096$, just below
$0.1$, while NBS's ($0.29$) and Jaccard's ($0.61$) stay large. ASE stays near $0.2$ under the school network's random rewiring, which lowers the sum
of squares of the four largest eigenvalues that its score tracks in every trial; it rises clearly above
$0.5$ only on the persistent school relabelling ($0.195$ to $0.810$), which keeps every eigenvalue. Under PAMAP2's
persistent relabelling, which moves $32\%$ of the edges, Hamming
has paired AUC $0.87$ [$0.86$, $0.88$].

\paragraph{Block models.} At matched degree the realized mean degree moves by at most $1.25\%$ across a
change. On its own generator LAD has EP $0.466$ [$0.337$, $0.589$] at its fixed setting ($0.587$ at matched degree) and
$0.991$ with the $499$ eigenvalues of its code. At matched degree in LAD's generator the counts have EP
$0.401$; LAD exceeds them by $+0.187$ [$-0.043$, $0.402$], Chen--Zhang by $+0.24$ [$0.02$, $0.46$], and
DeltaCon and the SpotLight-style sketch by $+0.068$ and $+0.122$ with intervals that cover zero. In s-GNN's
merge at matched degree SCPD exceeds the counts by $+0.162$ [$0.027$, $0.294$]. In s-GNN's exchange of $40$ vertex pairs the counts have EP $0.599$ [$0.489$, $0.705$]; Hamming and
Jaccard are $0.08$ and $0.12$ above them, not significantly.

\end{document}